\documentclass[prl,aps,twocolumn,floatfix,showpacs]{revtex4}
\usepackage{graphicx}
\usepackage{amsmath}
\usepackage{amssymb}
\usepackage{mathbbol}
\usepackage{times}
\usepackage[linktocpage]{hyperref}

\DeclareMathOperator{\im}{Im}
\DeclareMathOperator{\trace}{Tr}

\DeclareMathOperator{\I}{i}
\DeclareMathOperator{\E}{e}

\graphicspath{{.}}

\begin{document}

\title{Green-Function-Based Monte Carlo Method for Classical Fields Coupled
  to Fermions}

\author{Alexander Wei{\ss}e} 
\affiliation{Institut f{\"u}r Physik,
  Ernst-Moritz-Arndt-Universit{\"a}t Greifswald, 17487 Greifswald,
  Germany} 
\altaffiliation{New affiliation: Max-Planck-Institut f{\"u}r
  Mathematik, P.O.Box 7280, 53072 Bonn, Germany}

\date{January 7, 2009}

\begin{abstract}
  Microscopic models of classical degrees of freedom coupled to
  non-interacting fermions occur in many different contexts. Prominent
  examples from solid state physics are descriptions of colossal
  magnetoresistance manganites and diluted magnetic semiconductors, or
  auxiliary field methods for correlated electron systems. Monte Carlo
  simulations are vital for an understanding of such systems, but
  notorious for requiring the solution of the fermion problem with
  each change in the classical field configuration. We present an
  efficient, truncation-free $O(N)$ method on the basis of Chebyshev
  expanded local Green functions, which allows us to simulate systems
  of unprecedented size~$N$.
\end{abstract}

\maketitle


The numerical simulation of quantum lattice models is a key tool in
solid state research and many other fields of physics. One class of
problems, which is notoriously difficult to study, are fermions
coupled to classical degrees of freedom.  Such microscopic models can
arise if parts of a complex system are approximated classically. A
prominent example is the double-exchange model, which describes the
ferromagnetism of mixed-valence manganites on the basis of classical
$t_{2g}$-spins whose orientation affects the kinetic energy of $e_g$
valence electrons.~\cite{Ze51b,AH55,Ge60} 
Another example are Mn-doped (III,V) semiconductors, where itinerant
holes trigger a ferromagnetic ordering of the Mn spins.~\cite{SKM01}
A completely different route that leads to a coupling of fermions and
classical degrees of freedom are the auxiliary field methods, which
tackle the problem of interacting fermions. Here, a
Hubbard-Stratonovich transformation is used to decouple the two-body
interaction into non-interacting fermions in an auxiliary field, which
is summed over with Monte Carlo methods.~\cite{BSS81,HF86}

For all these systems the cause of the numerical difficulty is the
requirement for a solution of the non-interacting fermion problem
whenever the classical field is varied in a Monte Carlo simulation. We
propose an efficient local-update algorithm, which obtains the change
in fermionic energy directly from a few local Green functions. These
Green functions can easily be calculated by Chebyshev expansion,
superseding estimates of the density of states and thus trace
calculations.  We illustrate the efficiency of the approach with
simulations of the double-exchange model.


The models we consider in this work are of the general form
\begin{equation}\label{hamgen}
  H = \sum_{ij} c_i^{\dagger} A_{ij}(\vec\phi) c_j \,,
\end{equation}
where $c_i^{(\dagger)}$ are fermion creation (annihilation) operators
at lattice site $i$, and $\vec\phi$ is a classical field with one or
more components at each site. For example, the double-exchange
model~\cite{KA88,WLF01a} is given by the Hamiltonian
\begin{equation}\label{hamde}
  H = -\sum_{\langle ij\rangle} t_{ij} c_i^{\dagger}c_j\,,
\end{equation}
where the summation is over nearest-neighbor sites and the hopping
$t_{ij}$ depends on the orientation $\{\theta_i,\phi_i\}$ of classical
local spins at each site,
\begin{equation}\label{matde}
  t_{ij}^{} =
  \cos\tfrac{\theta_i - \theta_j}{2} \cos\tfrac{\phi_i - \phi_j}{2} +
  \I\cos\tfrac{\theta_i + \theta_j}{2} \sin\tfrac{\phi_i - \phi_j}{2}\,.
\end{equation}
This complex matrix element is one for ferromagnetically aligned spins
and vanishes for anti-ferromagnetic alignment. At low temperature the
system favors ferromagnetism, since it can gain kinetic energy.

The thermodynamics is described by the partition function
\begin{equation}
  Z = \trace_{\text{c}}\trace_{\text{f}}\exp[-\beta(H(\vec\phi)-\mu N)]
\end{equation}
and its derivatives. Here the traces $\trace_{\text{c}}$ and
$\trace_{\text{f}}$ sum over the classical and fermionic degrees of
freedom, respectively. The fermionic trace can be rewritten in terms of
the single-particle eigenvalues $\epsilon_i$ of $H$, 
\begin{equation}
  Z = \trace_{\text{c}} \exp[-S_{\text{eff}}(\vec \phi)]\,,
\end{equation}
such that the grand potential of the fermions times $\beta$,
\begin{equation}
  S_{\text{eff}}(\vec \phi) = 
  -\sum_i \log\big(1+\exp[-\beta(\epsilon_i(\vec\phi)-\mu)]\big)\,,
\end{equation}
defines an effective Euclidean action for the classical degrees of
freedom. The second trace over the classical field can then be
calculated with a standard Monte Carlo sampling, where the weight of a
configuration $\vec\phi$ is given by
\begin{equation}
  P(\vec\phi) = \exp[-S_{\text{eff}}(\vec \phi)]/Z\,.
\end{equation}
However, a non-trivial problem remains: To calculate $P(\vec\phi)$ we
need to know the spectrum $\{\epsilon_i\}$ of the non-interacting
fermion system, or at least its change under a proposed Monte
Carlo update $\vec\phi\to\vec\phi'$.

There are different solutions to this problem: We could use brute
force and calculate all eigenvalues of $A_{ij}(\vec\phi)$ whenever
$\vec\phi$ is modified. For local update schemes this is, of course,
very expensive and imposes severe restrictions on the accessible
system sizes. As a way out, hybrid
approaches~\cite{SSS86,DKPR87,AFGLM01,WFI05} have been suggested,
where updates of the whole field configuration are calculated with an
approximate dynamics. Then, the solution of the full fermion problem
in the acceptance step is required less frequently. However, if the
approximate dynamics does not closely match the exact one, the
acceptance rate drops markedly, in particular for increased system
size. Therefore, these approaches crucially depend on the quality of
the approximate action.

Staying with local updates of the classical field one can try to
optimize the calculation of the fermion spectrum. Motome and
Furukawa~\cite{MF99,MF00,MF01} suggested a Chebyshev expansion of the
fermion density of states, which can be calculated with an effort
proportional to the square of the system size $N$. A further
modification~\cite{FM04,ASFMD05}, involving several truncations in the
moment calculation, reduced the effort to order $N$.

In another recent approach~\cite{ANA07} the evolution of the
eigenvalues of $A_{ij}(\vec\phi)$ under small local changes of
$\vec\phi$ is tracked using special techniques for low-rank matrix
updates~\cite{GvL96}. Then again, the full solution of the fermion
problem is required only occasionally. In the best case this leads to
$N\log N$ scaling.

In the present work we combine ideas from the last two approaches
and directly calculate the change of the fermion density of states
using a few real space Green functions. Relying on Chebyshev expansion
these can be calculated with an effort proportional to the system size
$N$. Without any truncations we arrive at an order-$N$ algorithm.

Let us start from the Hamiltonian $H$ with the Hermitian hopping
matrix $A$ and ask how the spectrum changes, when a local modification
$\Delta$ is added. Given $a(E) = A - E\mathbb{1}$ and its inverse
$G(E)$, i.e. the Green function with
  $G(E)a(E) = \mathbb{1}$,
the spectrum of $A+\Delta$ follows from
\begin{equation}
  \begin{aligned}
    (A+\Delta)|\psi\rangle & = E|\psi\rangle\,,\\
    (a(E) + \Delta)|\psi\rangle & = 0\,.
  \end{aligned}
\end{equation}
Right multiplication with $G(E)$ yields
\begin{equation}\label{eq:first}
    G(E)(a(E) + \Delta)|\psi\rangle = 
    (\mathbb{1} + G(E)\Delta)|\psi\rangle = 0\,.
\end{equation}
The spectrum of $A+\Delta$ is given by those values of $E$ where the
determinant
\begin{equation}\label{eq:defdet}
  d(E) := \det(\mathbb{1} + G(E)\Delta)
\end{equation}
vanishes. Recalling introductory lectures on Green functions (see
e.g. Ref.~\onlinecite{Zi72}) we note that this $N$-dimensional
determinant reduces to one with a dimension equal to the rank of
$\Delta$. For local Monte Carlo updates this is a small number. If we
change, for instance, the on-site potential, $\Delta$ has a single
non-zero matrix element and we merely need the local Green function of
the corresponding site.  For the double exchange model a single spin
flip affects the hopping between the site and its nearest
neighbors. Independent of the system size $N$ or the space dimension,
Eq.~\eqref{eq:defdet} then reduces to a $2\times 2$ problem, i.e., we
need only four Green functions connecting the site with its
environment and both to themselves.

Before we go into the details of calculating $d(E)$, let us further
analyze the meaning of this quantity. Going back to Eq.~\eqref{eq:first} we
have 
\begin{equation}
  \begin{aligned}
    d(E) & = \det[G(E)(a(E) + \Delta)]\\
    & = \det[G(E)]\det[a(E) + \Delta]\,.
  \end{aligned}
\end{equation}
This can be expressed in terms of the eigenvalues $\{\epsilon_i\}$ of $A$ and
$\{\epsilon_i'\}$ of $A+\Delta$,
\begin{equation}
  d(E) = \prod_i \frac{1}{\epsilon_i -E}\prod_i (\epsilon_i'-E)\,.
\end{equation}
An important trick of our new approach consists of going to the complex
plane, $E\to z:=E+\I\varepsilon$, and taking a logarithmic derivative.
We now observe that $d(E)$ determines exactly what we need for the Monte
Carlo update: the change in the density of states going from $A$ to
$A+\Delta$, i.e., from $\vec\phi$ to $\vec\phi'$,
\begin{equation}
  \begin{aligned}
    \frac{1}{\pi} \im\lim_{\varepsilon\to 0} \frac{d \log(d(z))}{dz} 
    & = \frac{1}{\pi} \im\lim_{\varepsilon\to 0} 
    \sum_i \frac{1}{\epsilon_i-z} - \frac{1}{\epsilon_i'-z}\\
    & = \sum_i \delta(\epsilon_i-E) - \delta(\epsilon_i'-E)\\
    & = \rho(E)-\rho'(E)\,.
  \end{aligned}
\end{equation}
The change in the effective action then reads
\begin{equation}\label{eq:deltaseff}
  \begin{aligned}
    S_{\text{eff}}&(\vec \phi') - S_{\text{eff}}(\vec \phi)=\\
    & = \int \log(1+\E^{-\beta(E-\mu)})(\rho(E)-\rho'(E))dE\\
    & =\frac{\beta}{\pi} \int\frac{1}{1+\E^{\beta(E-\mu)}} 
    \im\lim_{\varepsilon\to 0}\log(d(E+\I\varepsilon)) dE\,,
  \end{aligned}
\end{equation}
where in the last line partial integration led to an integral over
the Fermi function. The key role of $d(z)$ has already been noted
earlier~\cite{FMPR81,SS81}, but the evaluation of
Eq.~\eqref{eq:defdet} becomes feasible only if we can restrict
ourselves to a minimal set of Green functions and an efficient,
direct method for their calculation.

\begin{figure}
  \centering
  \includegraphics[width=\linewidth]{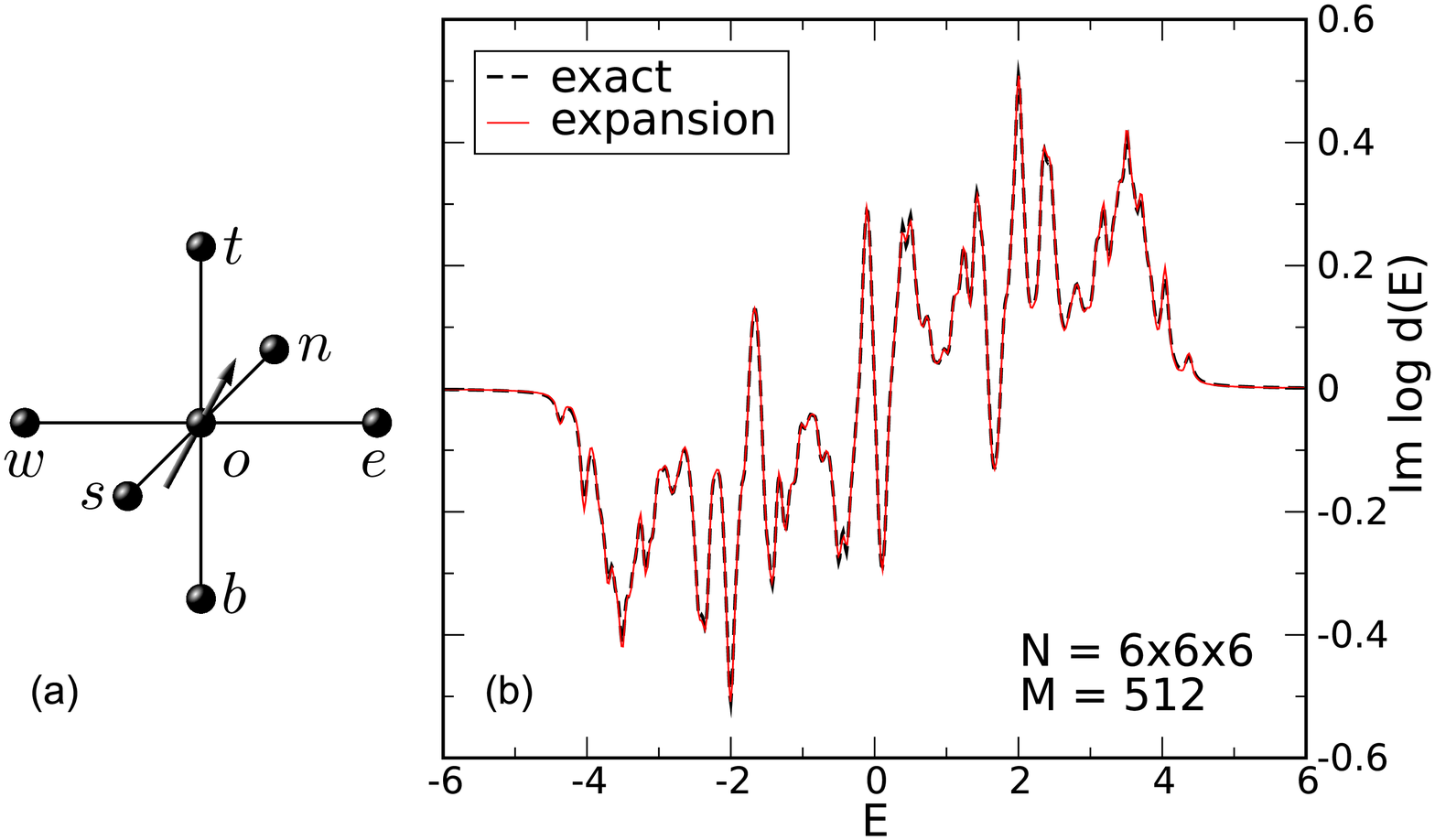}
  \caption{(color online). (a) Labeling of the nearest neighbor sites on a cubic
    lattice. (b) Comparison of $\im\log(d(E+\I\varepsilon))$
    calculated from exact eigenvalues $\epsilon_i$, $\epsilon_i'$ and
    with Chebyshev expansion of order $M=512$ ($\varepsilon=0.0625$) for
    a $6^3$ site sample.}
  \label{fig:nnlabels}
\end{figure}
Let us now explain this main part of our new approach for the double
exchange model on a cubic lattice. Here, a local update consists of
rotating a single spin at a site $o$. This modifies the matrix element
$t_{ij}$ between $o$ and its nearest neighbors to the north, east,
south, and so on. Labeling the sites according to
Fig.~\ref{fig:nnlabels}(a), a na\"ive evaluation of
Eq.~\eqref{eq:defdet} requires all $7\times 7$ Green functions
$G_{ij}(z)$ with $i,j\in\{o,n,e,s,w,t,b\}$~\cite{remark}. However, we
can do much better observing that
\begin{equation}
  \begin{aligned}
    d(z) & = \det(\mathbb{1} + G(z)\Delta)\\
    & = [1 + \sum_{j\in \text{n.n.}}\Delta_{jo} G_{oj}(z)]
    [1 + \sum_{j\in \text{n.n.}}\Delta_{oj} G_{jo}(z)]\\
    & - G_{oo}(z)[\sum_{j,k\in \text{n.n.}} \Delta_{jo}\Delta_{ok} G_{kj}(z)]
  \end{aligned}
\end{equation}
can be expressed in terms of only $2\times 2$ Green functions,
\begin{equation}\label{eq:simpledet}
  d(z) = [1 + G_{ov}(z)][1 + G_{vo}(z)] - G_{oo}(z)G_{vv}(z)\,,
\end{equation}
which connect the original site $o$ and the environment state
\begin{equation}
  |v\rangle = \Delta|o\rangle = \sum_{j\in \text{n.n.}} \Delta_{jo}|j\rangle\,.
\end{equation}

All four Green functions can be calculated easily with the Chebyshev
expansion approach outlined in a recent review~\cite{WWAF06}. In a
nutshell, diagonal elements $G_{ii}(z)$ are expanded in terms of the
Chebyshev polynomials of first and second kind, $T_m$ and $U_m$,
respectively,
\begin{multline}\label{eq:chebexp}
  G_{ii}(E+\I\varepsilon)  = 
  \frac{\I[\mu_0 + 2 \sum_{m=1}^{M-1} \mu_m T_m(E/s)]
  }{\sqrt{s^2-E^2}}\\
  + 2 \sum_{m=1}^{M-1} \mu_m U_{m-1}(E/s)\\
  = \I\frac{\mu_0 + 2 \sum_{m=1}^{M-1}\mu_m \exp[-\I m\arccos(E/s)]
  }{\sqrt{s^2-E^2}}\,.
\end{multline}
The expansion coefficients $\mu_m$ are Chebyshev moments modified by
appropriate kernel factors, which improve the convergence of the
truncated series and damp Gibbs oscillations,
\begin{equation}
  \mu_m = \langle i|T_m(H/s)|i\rangle 
  \frac{\sinh[\lambda (1-m/M)]}{\sinh\lambda}\,.
\end{equation}
The scaling factor $s$ ensures that the spectrum of the Hamiltonian
$H/s$ falls within the domain of the Chebyshev polynomials $[-1,1]$.
For the double exchange model we can choose $s$ to be a little larger
than the bare bandwidth in the ferromagnetic case, $s>6$.  The kernel
parameter $\lambda$ regulates the resolution of the method versus the
damping of Gibbs oscillations. A good value is $\lambda =4$. The
resolution at which the Green function is approximated is given by
$\varepsilon=\lambda s/M$. The limit $\varepsilon\to 0$ thus
corresponds to infinite expansion order $M$.  In a numerical
simulation, of course, $M$ is always finite and we need to extrapolate
data for different $M$ to obtain the limiting value of
$S_{\text{eff}}(\vec \phi') - S_{\text{eff}}(\vec \phi)$.  Since in
Eq.~\eqref{eq:deltaseff} we integrate a function of $G_{ij}(z)$ over
the Fermi function, the maximal resolution should be better than the
thermal broadening of the Fermi step, which is of the order of
$1/\beta$. Low temperatures, therefore, require higher expansion
orders.

The most time consuming step of the whole simulation is the
calculation of the moments $\langle i|T_m(H/s)|i\rangle$. Using the
recursion relation 
\begin{equation}\label{eq:chebrec}
  T_m(x) = 2 x T_{m-1}(x) - T_{m-2}(x)
\end{equation}
it reduces to sparse matrix-vector multiplications, the cost of which
scales linearly with the system size $N$. With a further trick based on
\begin{equation}
  T_{2m+i} = 2 T_{m} T_{m+i} - T_{i}
  \quad\text{with}\quad
  i=0,1
\end{equation}
we obtain two moments per matrix-vector multiplication. Moreover, half
of the moments vanish due to the relation $|v\rangle=\Delta|o\rangle$
and the special structure of the Chebyshev recursion~\eqref{eq:chebrec}.

At this point we should also note the advantage over approaches based
on a full expansion of the density of states: For the latter the
moments are given by traces, $\mu_m\sim\trace \{T_m(H/s)\}$, instead
of simple expectation values. Unless truncated or otherwise
approximated~\cite{FM04}, their calculation requires $O(N^2)$
operations.

\begin{figure}
  \centering
  \includegraphics[width=0.891\linewidth]{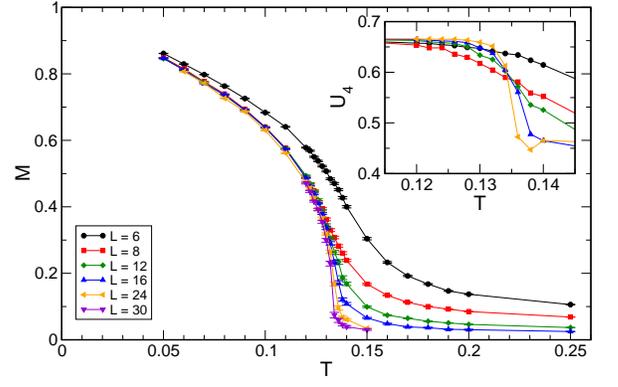}
  \caption{(color online). Main panel: Magnetization versus temperature for the
    double-exchange model on 3D clusters with periodic boundary
    conditions ($N = L^3$). Inset: Binder ratio $U_4$ showing a
    crossing near $T\approx 0.134$.}
  \label{fig:dedata}
\end{figure}

So far we have discussed only the diagonal Green functions
$G_{ii}$. In Ref.~\onlinecite{WWAF06} we showed that symmetric Green
functions $G_{ij} = G_{ji}$ can be expanded in the same
manner. However, in the double exchange model the spins induce local
magnetic fields which break this symmetry. We therefore derive the
off-diagonal Green functions $G_{ov}$ and $G_{vo}$ from the two
diagonal functions $G_{o+v,o+v}$ and $G_{o+\I v,o+\I v}$. With all
required moments at hand we can evaluate the sums in
Eq.~\eqref{eq:chebexp} with fast Fourier methods, calculate $d(E)$
with Eq.~\eqref{eq:simpledet}, and finally integrate over the Fermi
function to obtain the change of $S_{\text{eff}}$. In
Fig.~\ref{fig:nnlabels}(b) we show a typical example of
$\im\log(d(E+\I\varepsilon))$ and compare the expansion with the exact
result from a full diagonalization.


Having explained the technical details of the approach let us now
illustrate its applicability with a few results for the
double-exchange model~\eqref{hamde} at half filling, $\mu=0$. In the
main panel of Fig.~\ref{fig:dedata} we show the magnetization as a
function of temperature, where the latter is measured in units of the
maximal hopping amplitude $t\equiv 1$. As expected, we observe a
phase transition to a ferromagnetically ordered phase below $T\approx
0.14$. A closer inspection based on the Binder parameter~\cite{Bi81}
\begin{equation}
  U_4 = 1 - \frac{\langle m^4\rangle}{3\langle m^2\rangle^2}
\end{equation}
yields the estimate $T_c\approx 0.134$, see the inset of
Fig.~\ref{fig:dedata}. This agrees with previous
estimates~\cite{MF01,AFGLM01,MF03} of the critical temperature, which
range between 0.128 and 0.139. 

The data in Fig.~\ref{fig:dedata} is based on expansions of order
$M=256$ and averages over 6000 to 20000 Monte Carlo steps in the
critical region, where one step corresponds to $N$ spin flips.  It is
the low resource consumption which allows for these far more precise
calculations compared to previous studies~\cite{MF01,MF03}, which were based 
on Chebyshev expansions of order $M\approx 20$. Note also that we can handle
much larger systems with $N=30^3$ sites, i.e. a complex matrix dimension
of $27000$. 

\begin{figure}
  \centering
  \includegraphics[width=\linewidth]{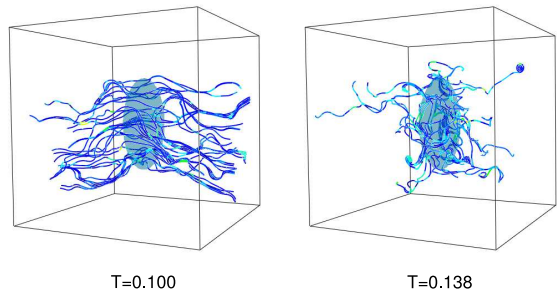}
  \caption{(color online). Structure of the 3D spin field at two different
    temperatures visualized with stream lines.}
  \label{fig:flow}
\end{figure}
To visualize the spatial structure of the spin field we imagine it as
a liquid flow and draw curves tangent to the velocity field.  These
stream lines are regular and parallel in the ordered phase, but quite
irregular and swirled near and above the phase transition, see
Fig.~\ref{fig:flow}. Of course, this visualization method fails for
truly disordered spin fields.

In summary, we have presented an efficient local update scheme for
Monte Carlo simulations of classical fields coupled to fermions. At the
core of the approach is an expression which relates the change of the
fermion spectrum to a few local Green functions. These can be
calculated easily with Chebyshev expansion. Compared to similar
expansion approaches a full trace over the fermion system is spared,
which directly leads to a fast and precise order-$N$ algorithm.
Possibly our method can be further accelerated using the
approximations inherent in the truncated polynomial expansion
method~\cite{FM04,ASFMD05}. The calculations presented were performed on the 
TeraFLOPS cluster of the Institute for Physics at Greifswald university.


\begin{thebibliography}{26}
\bibitem[{rem()}]{remark}
\bibinfo{note}{In Ref.~\onlinecite{SS81} this is described as a reduction to a
  $L\times L$ problem.}
\end{thebibliography}

\end{document}